\documentstyle[aps,epsfig]{revtex}
\begin{document}
\title{The sound velocity in an equilibrium hadron gas
\thanks{ Work partially supported by the Polish Committee for 
Scientific
Research under contract KBN - 2 P03B 030 18}}
\author{Dariusz Prorok and Ludwik Turko}
\address{Institute of Theoretical Physics, University of
Wroc{\l}aw,\\ Pl.Maksa Borna 9, 50-204  Wroc{\l}aw, Poland}
\date{January 23, 2001}
\maketitle
\begin{abstract}
We calculate the velocity of sound in an ideal gas of massive
hadrons with non-vanishing baryon number. The gas is in thermal
and chemical equilibrium. Also we show that the temperature
dependence $T(\tau) \cong T_{0} \cdot \left( {\tau_{0} \over \tau}
\right)^{c_{s}^{2}}$ is approximately valid, when the gas expands
longitudinally according to the Bjorken law.
\end{abstract}

\section {Introduction }

The main subject of the following paper is the calculation of the
sound velocity $c_{s}$ in the multi-component hadron gas with
nonzero baryon number density. The sound velocity appears in the
hydrodynamical description of the matter which is created in the
central rapidity region (CRR) of a heavy-ion collision. It turned
out \cite{Bjorken:1983qr,Baym:1983sr} that the evolution of the
matter (whatever it is: a quark-gluon plasma or the hadron gas)
proceeds as longitudinal and superimposed transverse expansions.
And the latter has the form of the rarefaction wave moving
radially inward with the velocity of sound. The second place where
the sound velocity appears is the temperature equation for an
ideal baryonless gas (it can be a gas of quarks and gluons as
well), which cools as the result only of the longitudinal
expansion (see e.g. \cite{Cleymans:1986wb})

\begin{equation}
T(\tau) = T_{0} \cdot \left( {\tau_{0} \over \tau}
\right)^{c_{s}^{2}} \;\;, \label{1}
\end{equation}

\noindent where $\tau$ is a local proper time, $\tau_{0}$ an
initial moment of time and $c_{s}$ is assumed constant. In the
following, we show that the above equation, now as some
approximation, is also valid in some range of temperature for the
hadron gas with non-vanishing baryon number.

In our previous paper \cite{Prorok:2000kv} we showed that taking
into account also the transverse expansion (in the the form of the
rarefaction wave) changes the $J/\Psi$ theoretical suppression
pattern qualitatively. This was done for the scenario without the
quark-gluon plasma appearance in the CRR. Of course, this is very
important because $J/\Psi$ suppression observed in NA38 
\cite{Abr}
and NA50 \cite{Abr50} heavy-ion collision data is treated as the
main signature of the creation of the quark-gluon plasma (QGP)
during these collisions \cite{Matsui:1986dk}. Since the study of
$J/\Psi$ suppression patterns was the main purpose of our paper
\cite{Prorok:2000kv}, we left aside the more complete presentation
of the behaviour of the sound velocity in the hadron gas. Now, we
would like to present our results in the form of a separate paper.

\section { The expanding multi-component hadron gas }
\label{hadgas}

For an ideal hadron gas in thermal and chemical equilibrium, which
consists of $l$ species of particles, energy density $\epsilon$,
baryon number density $n_{B}$, strangeness density $n_{S}$,
entropy density $s$ and pressure $P$ read ($\hbar=c=1$ always)

\begin{mathletters}
\label{eqstate}
\begin{equation}
\epsilon = { 1 \over {2\pi^{2}}} \sum_{i=1}^{l} (2s_{i}+1)
\int_{0}^{\infty} { { dpp^{2}E_{i} } \over { \exp \left\{ {{ E_{i}
- \mu_{i} } \over T} \right\} + g_{i} } } \ , \label{2a}
\end{equation}

\begin{equation}
n_{B}={ 1 \over {2\pi^{2}}} \sum_{i=1}^{l} (2s_{i}+1)
\int_{0}^{\infty} { { dpp^{2}B_{i} } \over { \exp \left\{ {{ E_{i}
- \mu_{i} } \over T} \right\} + g_{i} } } \ , \label{2b}
\end{equation}

\begin{equation}
n_{S}={1 \over {2\pi^{2}}} \sum_{i=1}^{l} (2s_{i}+1)
\int_{0}^{\infty} { { dpp^{2}S_{i} } \over { \exp \left\{ {{ E_{i}
- \mu_{i} } \over T} \right\} + g_{i} } } \ , \label{2c}
\end{equation}

\begin{equation}
s={1 \over {6\pi^{2}T^{2}} } \sum_{i=1}^{l} (2s_{i}+1)
\int_{0}^{\infty} { {dpp^{4}} \over { E_{i} } } { { (E_{i} -
\mu_{i}) \exp \left\{ {{ E_{i} - \mu_{i} } \over T} \right\} }
\over { \left( \exp \left\{ {{ E_{i} - \mu_{i} } \over T} \right\}
+ g_{i} \right)^{2} } }\ , \label{2d}
\end{equation}

\begin{equation}
P={1 \over {6\pi^{2}} } \sum_{i=1}^{l} (2s_{i}+1)
\int_{0}^{\infty} { {dpp^{4}} \over { E_{i} } } { 1 \over { \exp
\left\{ {{ E_{i} - \mu_{i} } \over T} \right\} + g_{i} } }\ ,
\label{2e}
\end{equation}
\end{mathletters}

\noindent where $E_{i}= ( m_{i}^{2} + p^{2} )^{1/2}$ and $m_{i}$,
$B_{i}$, $S_{i}$, $\mu_{i}$, $s_{i}$ and $g_{i}$ are the mass,
baryon number, strangeness, chemical potential, spin and a
statistical factor of specie $i$ respectively (we treat an
antiparticle as a different specie). And $\mu_{i} = B_{i}\mu_{B} +
S_{i}\mu_{S}$, where $\mu_{B}$ and $\mu_{S}$ are overall baryon
number and strangeness chemical potentials respectively.

We shall work here within the usual timetable of the events in the
CRR of a given ion collision (for more details see e.g.
\cite{Blaizot:1989ec}). We fix $t=0$ at the moment of the maximal
overlap of the nuclei and assume that it takes place at $z=0$,
where $z$ is the coordinate of a collision axis. After half of the
time the nuclei need to cross each other, matter appears in the
CRR. We assume that soon thereafter matter thermalizes and this
moment, $\tau_{0}$ ($\tau=\sqrt{t^{2}-z^{2}}$), is estimated at
about 1 fm \cite{Blaizot:1989ec,Bjorken:1983qr}. Then matter
starts to expand and cool and after reaching the freeze-out
temperature $T_{f.o.}$ it is no longer a thermodynamical system.
As we have already mentioned in the introduction, we assume that
this matter is the hadron gas, which consists of all hadrons up to
$\Omega^{-}$ baryon. The expansion proceeds according to the
relativistic hydrodynamics equations and for the longitudinal
component we have the following solution (for details see e.g.
\cite{Bjorken:1983qr,Cleymans:1986wb})

\begin{equation}
s(\tau)= { {s_{0}\tau_{0}} \over {\tau} } \;,\;\;\;\;\;\;\;\;\;
n_{B}(\tau)= { {n_{B}^{0}\tau_{0}} \over {\tau} } \ , \label{3}
\end{equation}

\noindent where $s_{0}$ and $n_{B}^{0}$ are initial densities of
the entropy and the baryon number respectively. The transverse
expansion has the form of the rarefaction wave moving radially
inward with a sound velocity $c_{s}$
\cite{Bjorken:1983qr,Baym:1983sr}.

To obtain the time dependence of temperature and baryon number 
and
strangeness chemical potentials one has to solve numerically
equations (\ref{2b} - \ref{2d})  with $s$, $n_{B}$ and $n_{S}$
given as time dependent quantities. For $s(\tau)$, $n_{B}(\tau)$
we have expressions (\ref{3}) and $n_{S}=0$ since we put the
overall strangeness equal to zero during all the evolution (for
more details see \cite{Prorok:1995xz}).

\section {The sound velocity in the multi-component hadron gas}
\label{soundv}

In the hadron gas the sound velocity squared is given by the
standard expression

\begin{equation}
c_{s}^{2}= { {\partial P} \over {\partial \epsilon} }\ . \label{4}
\end{equation}

\noindent Since the experimental data for heavy-ion collisions
suggests that the baryon number density is non-zero in the CRR at
AGS and SPS energies
\cite{Baechler:1991pd,Stachel:1999rc,Ahle:1998jc}, we calculate
the above derivative for various values of $n_{B}$.

\begin{figure}
\begin{center}{
{\epsfig{file=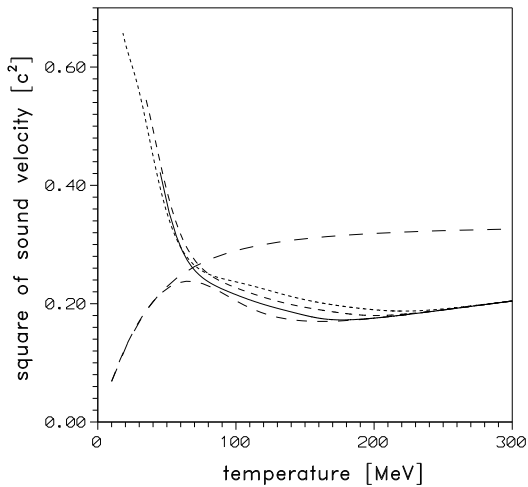,width=7cm}} }\end{center}
\caption{Dependence of the sound velocity squared on temperature
for various values of $n_{B}$: $n_{B}=0.65\;fm^{-3}$
(short-dashed), $n_{B}= 0.25\;fm^{-3}$ (dashed),
$n_{B}=0.05\;fm^{-3}$ (solid) and $n_{B}=0$ (long-dashed). The
case of the pure pion gas (long-long-dashed) is also presented.}
\label{Fig.1.}
\end{figure}

To estimate initial baryon number density $n_{B}^{0}$ we can use
experimental results for S-S \cite{Baechler:1991pd} or Au-Au
\cite{Stachel:1999rc,Ahle:1998jc} collisions. In the first
approximation we can assume that the baryon multiplicity per unit
rapidity in the CRR is proportional to the number of participating
nucleons. For a sulphur-sulphur collision we have $dN_{B}/dy \cong
6$ \cite{Baechler:1991pd} and 64 participating nucleons. For the
central collision of lead nuclei we can estimate the number of
participating nucleons at $2A = 416$, so we have $dN_{B}/dy \cong
39$. Having taken the initial volume in the CRR equal to $\pi
R_{A}^{2} \cdot 1$ fm, we arrive at $n_{B}^{0} \cong 0.25 \;
fm^{-3}$. This is some underestimation because the S-S collisions
were at a beam energy of $200\; GeV$/nucleon, but Pb-Pb at
$158\;GeV$/nucleon. From the Au-Au data extrapolation one can
estimate $n_{B}^{0} \cong 0.65 \; fm^{-3}$ \cite{Stachel:1999rc}.
These values are for central collisions. So, we estimate (\ref{4})
for $n_{B}= 0.25,\;0.65\;fm^{-3}$ and additionally, to investigate
the dependence on $n_{B}$ much carefully, for
$n_{B}=0.05\;fm^{-3}$. The results of numerical evaluations of
(\ref{4}) are presented in Fig.~ \ref{Fig.1.}. For comparison, we
drew also two additional curves: for $n_{B}=0$ and for a pure
massive pion gas. These curves are taken from
\cite{Prorok:1995xz}.

The "physical region" lies between more or less $100$ and
$200\;MeV$ on the temperature axis. This is because the critical
temperature for the possible QGP-hadronic matter transition is of
the order of $200\;MeV$ \cite{Ukawa:1998vz} and the freeze-out
temperature should not be lower than $100\;MeV$
\cite{Stachel:1999rc}. In low temperatures we can see completely
different behaviours of cases with $n_{B}=0$ and $n_{B}\not=0$. 
We
think that this is caused by the fact that for $n_{B}\not=0$ the
gas density can not reach zero when $T\rightarrow 0$, whereas for
$n_{B}=0$ it can. For the higher temperatures all curves excluding
the pion case behave in the same way qualitatively. From $T
\approx 70\;MeV$ they decrease to their minima (for
$n_{B}=0.65\;fm^{-3}$ at $T \cong 221.3\;MeV$, for
$n_{B}=0.25\;fm^{-3}$ at $T \cong 201.5\;MeV$, for
$n_{B}=0.05\;fm^{-3}$ at $T \cong 176.6\;MeV$ and for $n_{B}=0$ 
at
$T \cong 160\;MeV$) and then they increase to cover each other
above $T \approx 250\;MeV$. To investigate this problem we check
the content of the hadron gas when the temperature changes.

\begin{figure}[htb]
\begin{minipage}[t]{75mm}
{\psfig{figure= 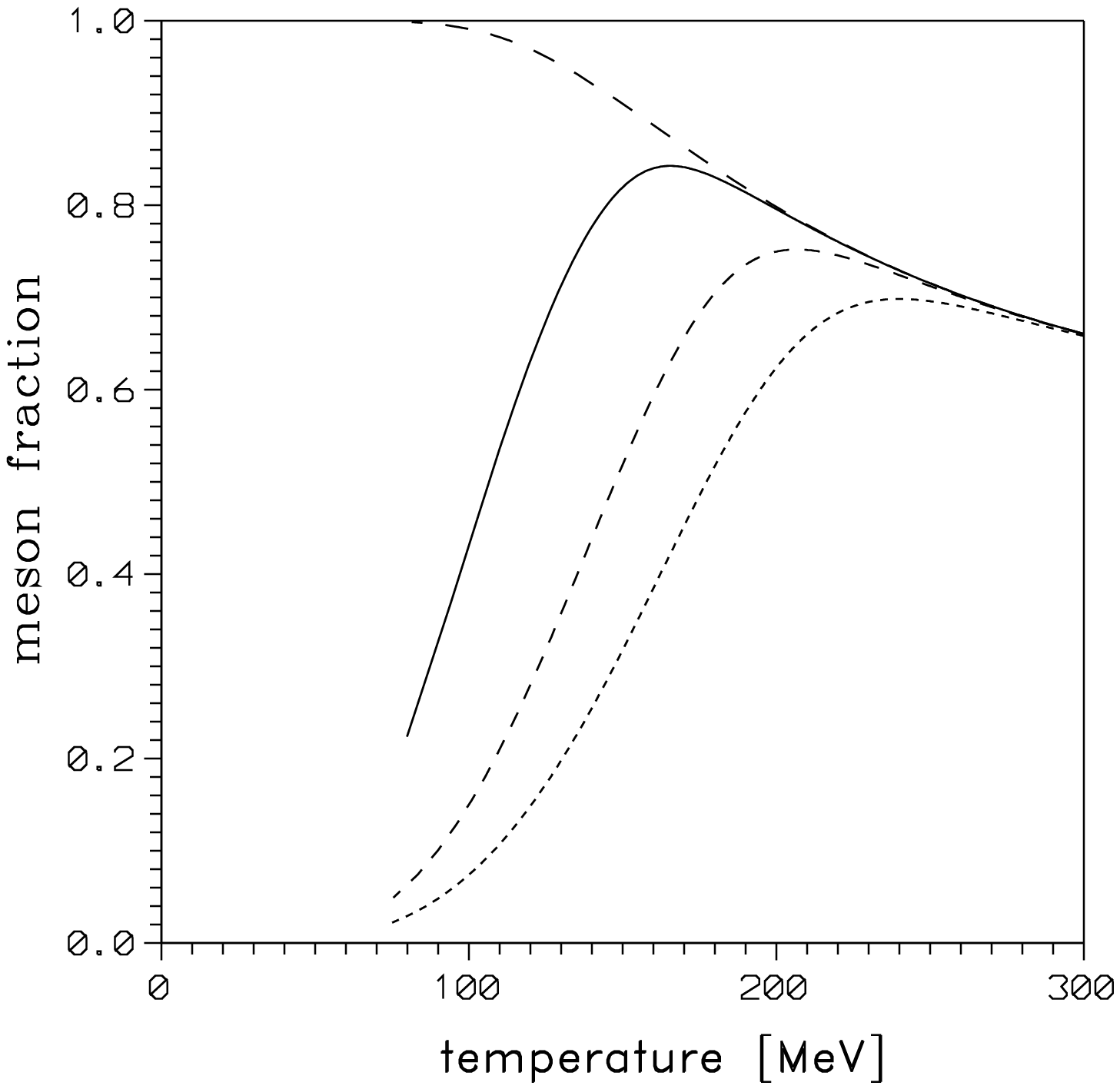,height=\textwidth,width=\textwidth}}
\caption{Fraction of mesons in the multi-component hadron gas as 
a
function of temperature for $n_{B}=0.65\;fm^{-3}$ (short-dashed),
$n_{B}= 0.25\;fm^{-3}$ (dashed), $n_{B}=0.05\;fm^{-3}$ (solid) and
$n_{B}=0$ (long-dashed).} \label{Fig.2.}
\end{minipage}
\hspace{\fill}
\begin{minipage}[t]{75mm}
{\psfig{figure= 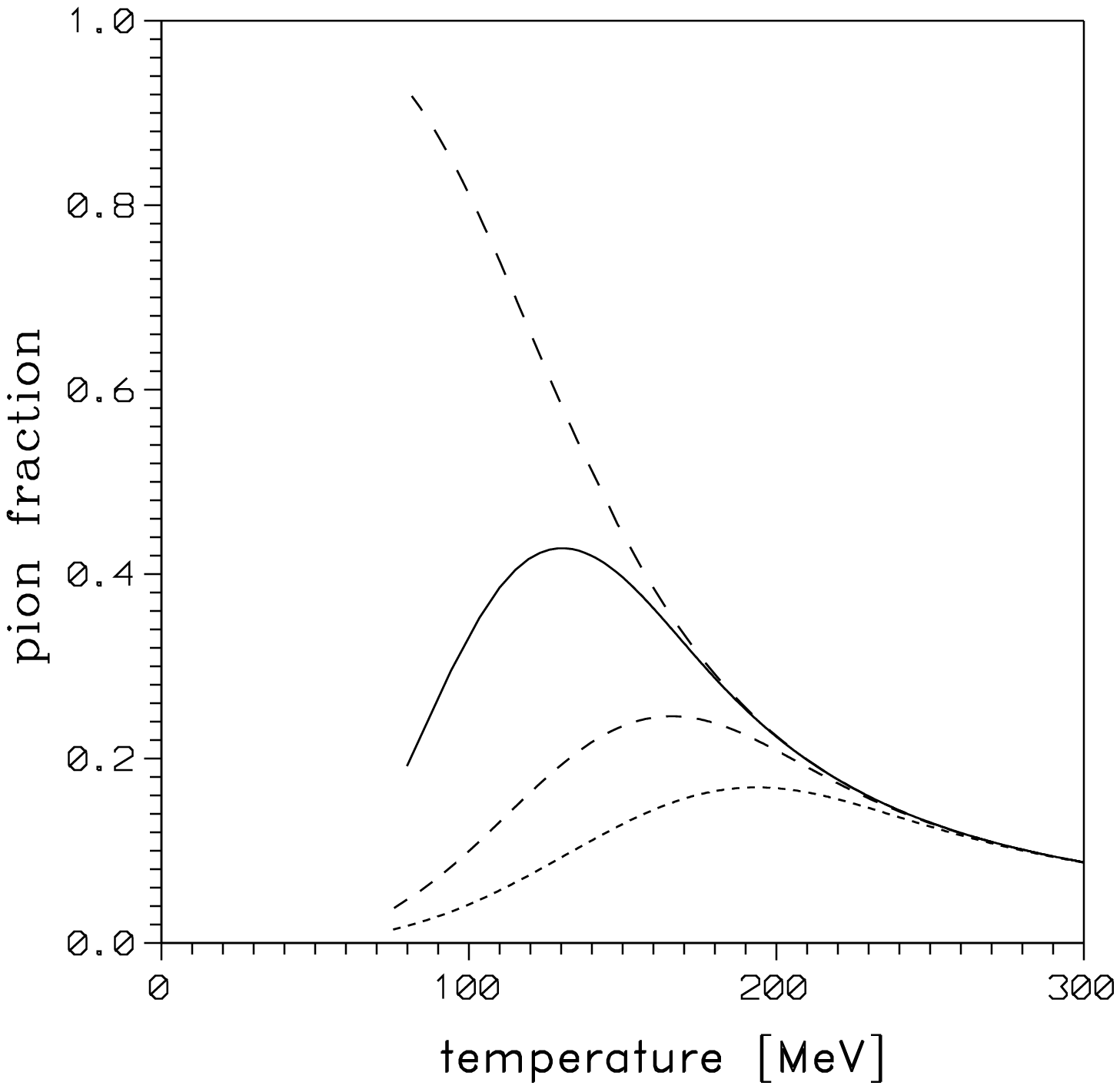 ,height=\textwidth,width=\textwidth}}
\caption{Fraction of pions in the multi-component hadron gas as a
function of temperature for $n_{B}=0.65\;fm^{-3}$ (short-dashed),
$n_{B}= 0.25\;fm^{-3}$ (dashed), $n_{B}=0.05\;fm^{-3}$ (solid) and
$n_{B}=0$ (long-dashed).} \label{Fig.3.}
\end{minipage}
\end{figure}

\begin{figure}[htb]
\begin{minipage}[t]{75mm}
{\psfig{figure= 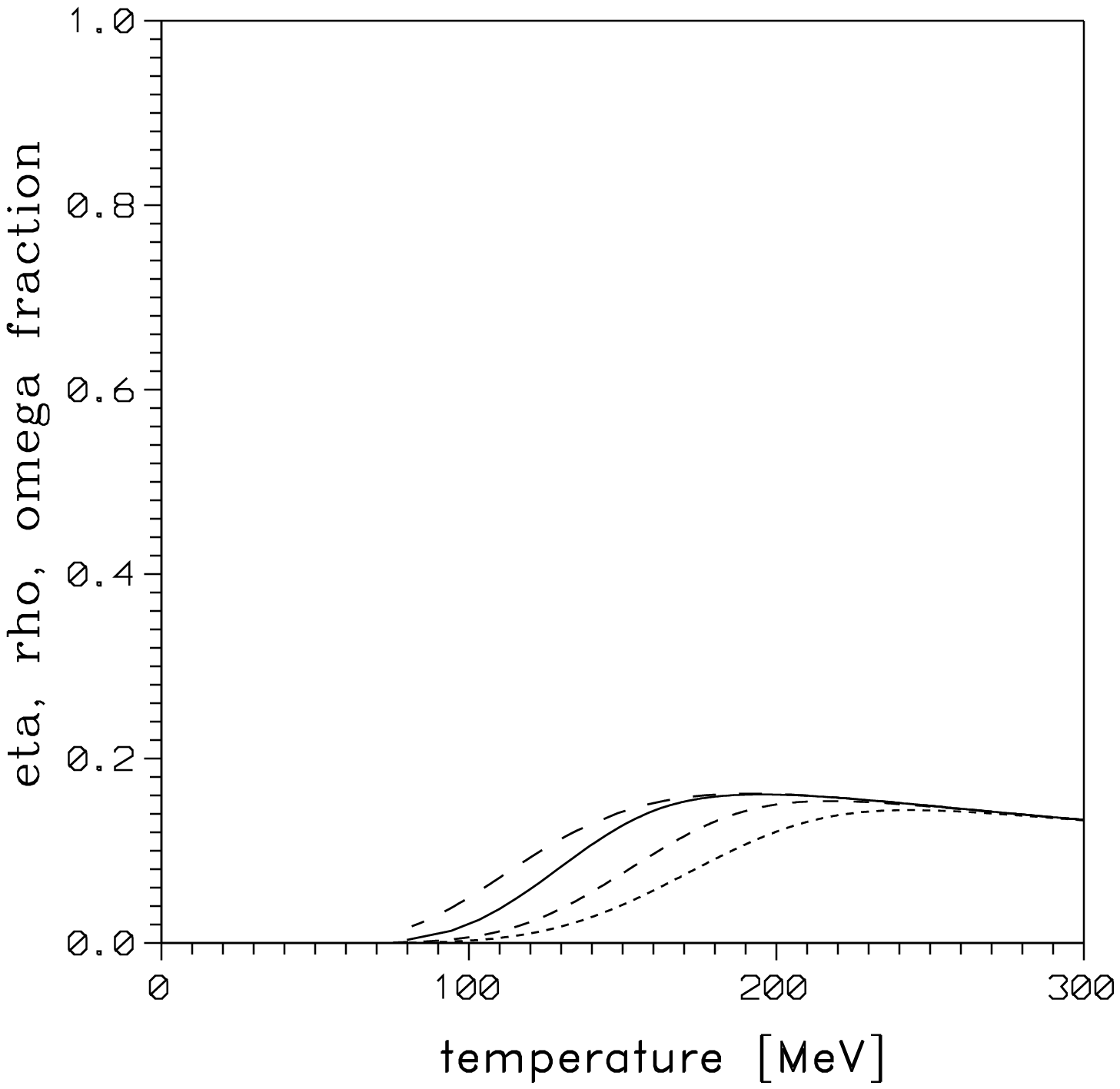,height=\textwidth,width=\textwidth}}
\caption{Fraction of $\eta$, $\rho$ and $\omega$ mesons in the
multi-component hadron gas as a function of temperature for
$n_{B}=0.65\;fm^{-3}$ (short-dashed), $n_{B}= 0.25\;fm^{-3}$
(dashed), $n_{B}=0.05\;fm^{-3}$ (solid) and $n_{B}=0$
(long-dashed).} \label{Fig.4.}
\end{minipage}
\hspace{\fill}
\begin{minipage}[t]{75mm}
{\psfig{figure= 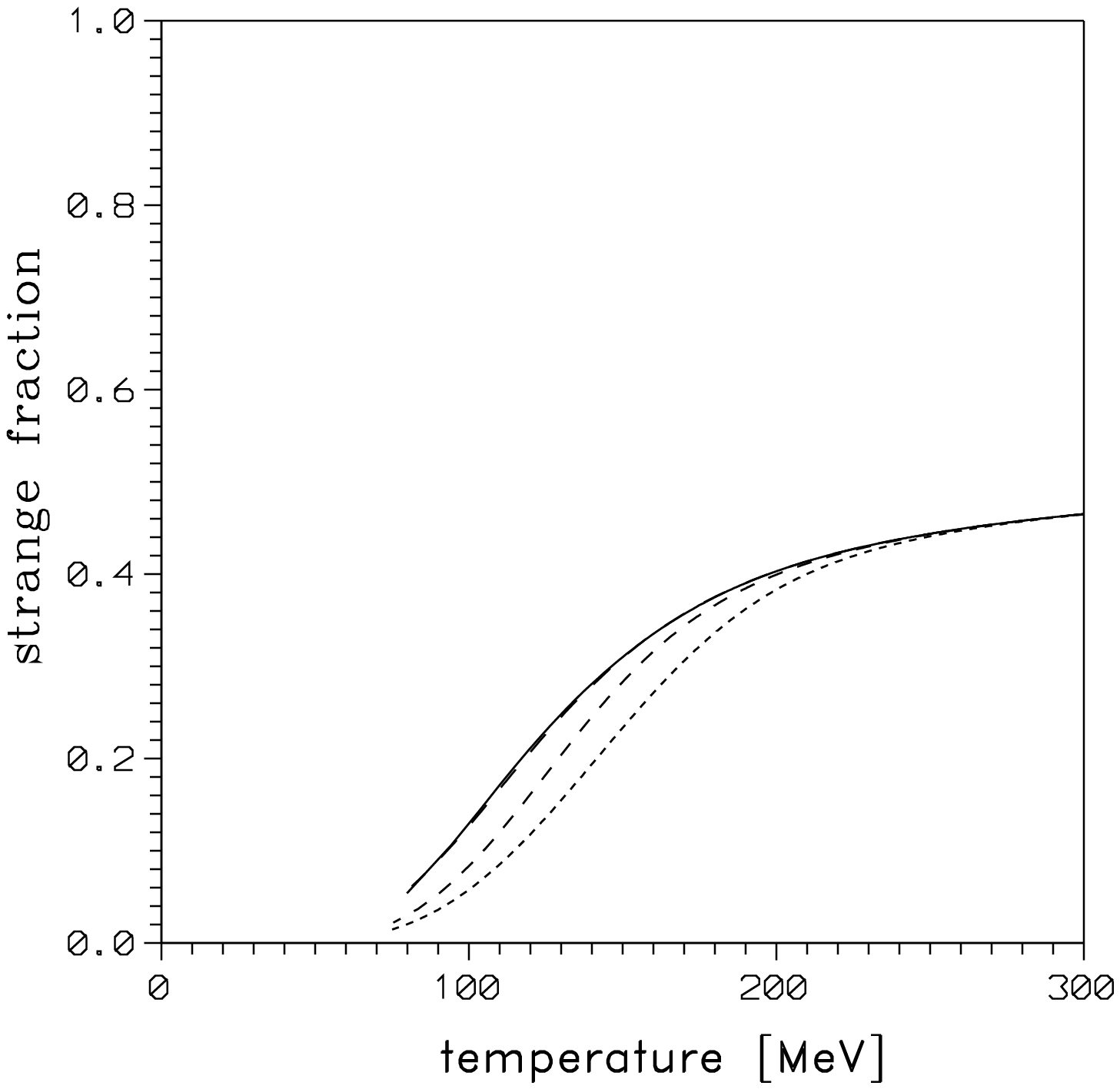 ,height=\textwidth,width=\textwidth}}
\caption{Fraction of strange particles in the multi-component
hadron gas as a function of temperature for $n_{B}=0.65\;fm^{-3}$
(short-dashed), $n_{B}= 0.25\;fm^{-3}$ (dashed),
$n_{B}=0.05\;fm^{-3}$ (solid). The curve for $n_{B}=0$
(long-dashed) covers the curve for $n_{B}=0.05\;fm^{-3}$ almost
completely.} \label{Fig.5.}
\end{minipage}
\end{figure}

We define the fraction of some kind of particles in the hadron gas
as

\begin{equation}
f_{p} = { \varrho_{p} \over \varrho }\ , \label{5}
\end{equation}

\noindent where $\varrho_{p}$ is the density of these particles
and $\varrho$ the density of the hadron gas. We calculate the
above-mentioned fractions of: a) mesons; b) pions; c) $\eta$,
$\rho$ and $\omega$ mesons and d) strange particles. The results
are presented in Figs.~\ref{Fig.2.}-\ref{Fig.5.}.

Coming back to the behaviour of the sound velocity of the hadron
gas above $T \approx 70\;MeV$ (see Fig.~ \ref{Fig.1.}), we can
state that the appearance of heavier particles, especially $\eta$,
$\rho$ and $\omega$ mesons, is responsible for it. First of all,
the curve for the pure pion gas in Fig.~ \ref{Fig.1.} is
increasing everywhere, so we have to take into account also other
species to obtain the sound velocity decreasing in some range of
temperature above $T \approx 70\;MeV$. If we assume that baryons
cause the mentioned behaviour of the sound velocity, it would not
agree with the case of $n_{B}=0$, because the baryon fraction is
increasing with the temperature everywhere for this case, see
Fig.~ \ref{Fig.2.} up-side-down. The same would happen for strange
particles (with the only difference that the strange particle
fraction is increasing with the temperature for all cases of
$n_{B}$, see Fig.~ \ref{Fig.5.}). Only the fraction of $\eta$,
$\rho$ and $\omega$ mesons changes its monotonicity above $T
\approx 70\;MeV$ and at the temperatures similar to those where
the sound velocity curves have their local minima (cf. Fig.~
\ref{Fig.1.} and Fig.~ \ref{Fig.4.}), for all cases of $n_{B}$.

\section {The pattern of cooling and its connection with the sound 
velocity}
\label{power}

In Sect.~\ref{hadgas} we have explained how to obtain the time
dependence of the temperature of the longitudinally expanding
hadron gas. This dependence proved to be very well approximated 
by
the expression

\begin{equation}
T(\tau) \cong T_{0} \cdot \tau^{-a}\ .  \label{6}
\end{equation}

\begin{figure}
\begin{center}{
{\epsfig{file=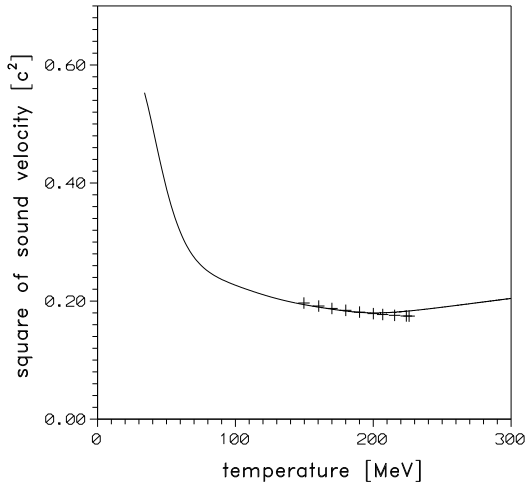,width=7cm}} }\end{center} \caption{The
power $a$ (crosses) from the approximation (\ref{6}) and the sound
velocity curve for $n_{B}= 0.25\;fm^{-3}$ .} \label{Fig.6.}
\end{figure}

\noindent The above approximation is valid in the temperature
ranges $[T_{f.o.},T_{0}]$, where $T_{f.o.} \geq 100\;MeV$, $T_{0}
\leq T_{0,max}$ and $T_{0,max} \approx 225\;MeV$. We started 
from
$T_{0}$ equal to $221.8\;MeV$ (for $n_{B}^{0} = 0.65\;fm^{-3}$),
$226\;MeV$ (for $n_{B}^{0} = 0.25\;fm^{-3}$) and $226.7\;MeV$ (for
$n_{B}^{0} = 0.05\;fm^{-3}$). These values correspond to
$\epsilon_{0} = 3.7\;GeV/fm^{3}$, which is the initial energy
density in the CRR slightly above the value $\epsilon_{0} =
3.5\;GeV/fm^{3}$ estimated by NA50 \cite{Abr50}. Then we took
several decreasing values of $T_{0} < T_{0,max}$. For every
$T_{0}$ chosen we repeat the procedure of obtaining the
approximation (\ref{6}), i.e. the power $a$. The values of $a$ as
a function of $T_{0}$ are depicted in Fig.~ \ref{Fig.6.} for the
case of $n_{B}= 0.25\;fm^{-3}$, together with the corresponding
sound velocity curve.

We can see that in the above-mentioned interval of temperature the
power $a$ has the meaning of the sound velocity within quite well
accuracy. We would like to add that we arrived at the same
conclusion also for cases with $n_{B}=0.65\;fm^{-3}$ and
$n_{B}=0.05\;fm^{-3}$. Note that the higher temperature, the worse
accuracy. This results from the fact that for a higher temperature
the wider interval is used to obtain the formula (\ref{6}).
Therefore the approximation is worse.

We can formulate the following conclusion: in the "physical
region" of temperature and for realistic baryon number densities,
the longitudinal expansion given by (\ref{3}) results in the
cooling of the hadron gas described by (\ref{6}) with
$a=c_{s}^2(T_{0})$. Note that $T(\tau) = T_{0} \cdot \left(
{\tau_{0} \over \tau} \right)^{c_{s}^{2}}$ is the exact expression
for a baryonless gas with the sound velocity constant (for details
see \cite{Baym:1983sr,Cleymans:1986wb}).

\section{Conclusions}
\label{conc}

In this paper we have calculated the sound velocity in the
multi-component hadron gas with non-zero baryon number density.
Such a gas, instead of the QGP,  could have appeared in the CRR 
of
heavy-ion collisions at AGS and SPS energies. We have compared 
the
results with the sound velocities in a gas with $n_{B}= 0$ and in
a pion gas. We have shown that in the "physical region"
(temperatures between roughly $100$ and $200\;MeV$) the sound
velocities for various cases of the multi-component hadron gas
behave qualitatively the same: firstly they decrease and then they
start to increase with the temperature. Comparison with the case
of the pion gas and analysis of fractions of particles in the
multi-component hadron gas suggest that the appearance of $\eta$,
$\rho$, and $\omega$ mesons is responsible for that.

The second result shown here is that for the multi-component
hadron gas the cooling imposed by the longitudinal expansion can
be described by the approximation

\begin{equation}
T(\tau) \cong T_{0} \cdot \left( {\tau_{0} \over \tau}
\right)^{c_{s}^{2}(T_{0})}\, \label{7}
\end{equation}

\noindent where $T_{0}$ belongs to "physical region". The
approximation is valid for the gas with the non-zero baryon number
density. The same conclusion was drawn for the case with 
$n_{B}=0$
in \cite{Prorok:1995xz}.

The values of the sound velocity obtained in the "physical region"
and the validity of approximation (\ref{7}) therein, cause that
the cooling of the hadron gas is much slower than the cooling of
an ideal massless gas with $n_{B}=0$ (where $c_{s}^{2}= {1 \over
3}$). This fact has the straightforward consequence for $J/\Psi$
suppression: the longer hadrons last as a gas, the deeper
suppression takes place (for details see \cite{Prorok:2000kv}).

\end{document}